# Voltage control of the long-range *p-d* exchange coupling in a ferromagnet-semiconductor quantum well hybrid structure

V.L. Korenev[1,*], I.V. Kalitukha[1], I.A. Akimov[1,2,*], V.F. Sapega[1], E.A. Zhukov[1,2], E. Kirstein[2], O.S. Ken[1], D. Kudlacik[2], G. Karczewski[3], M. Wiater[4], T. Wojtowicz[4], N.D. Ilyinskaya[1], N.M. Lebedeva[1], T.A. Komissarova[1], Yu.G. Kusrayev[1], D.R. Yakovlev[1,2], and M. Bayer[1,2]

[1]Ioffe Institute, Russian Academy of Sciences, 194021 St. Petersburg, Russia
[2]Experimentelle Physik 2, Technische Universität Dortmund, D-44227 Dortmund, Germany
[3]Institute of Physics, Polish Academy of Sciences, PL-02668 Warsaw, Poland
[4]International Research Centre MagTop, Institute of Physics, Polish Academy of Sciences
PL-02668 Warsaw, Poland

*Voltage control of ferromagnetism on the nanometer scale is highly appealing for the development of novel electronic devices. Here a key challenge is to implement and combine low power consumption, high operation speed, reliable reversibility and compatibility with semiconductor technology. Hybrid structures based on the assembly of ferromagnetic and semiconducting building blocks are attractive candidates in that respect as such systems bring together the properties of the isolated constituents: They are expected to show magnetic order as a ferromagnet and to be electrically tunable as a semiconductor. Here we demonstrate the electrical control of the exchange coupling in a hybrid consisting of a ferromagnetic Co layer and a semiconductor CdTe quantum well, separated by a thin non-magnetic (Cd,Mg)Te barrier. The effective magnetic field of the exchange interaction reaches up to 2.5 Tesla and can be turned on and off by application of 1 V bias across the heterostructure. The mechanism of this*

*electric field control is essentially different from the conventional concept, in which wavefunctions are spatially redistributed to vary the exchange interaction, requiring high field strengths. Here we address instead control of the novel exchange mechanism that is mediated by elliptically polarized phonons emitted from the ferromagnet, i.e. the phononic ac Stark effect. An essential parameter of this coupling is the splitting between heavy and light hole states in the quantum well which can be varied by the electric field induced band bending. Thereby the splitting can be tuned with respect to the magnon-phonon resonance energy in the ferromagnet, leading to maximum coupling for flat band conditions. Our results demonstrate the feasibility of electrically controlled exchange coupling in hybrid semiconductor nanostructures at quite moderate electric field strengths.*

*Corresponding author: korenev.orient@mail.ioffe.ru, ilja.akimov@tu-dortmund.de

**Introduction**

Nowadays the demand for control of ferromagnetism on the nanometer scale is met by the methods of spin-transfer torque or spin-orbit torque, both based on locally controlled magnetization reversal by a high-density current of ~$10^6$ A/cm$^2$ [1]. However, more promising in terms of energy costs is the use of an electric field, instead of electrical current or magnetic field, which would allow fast voltage control of magnetism [2]. This type of control was realized, for instance, for the low-temperature magnetic semiconductors (In,Mn)As and (Ga,Mn)As [3], and more recently significant progress in that direction was achieved in various materials. Examples are the coercive force in multiferroics [4], the magnetic anisotropy in ultrathin Fe/MgO [2] and the magnetic order in a ferromagnet-ferroelectric structure [5]. The most intriguing idea for tuning magnetic properties is based on the control of the exchange interaction causing the magnetism (the strongest spin-spin interaction) through varying the carrier wavefunction overlap in a thin magnetic layer. However, this requires application of rather strong electric fields of ~$10^7$ V/cm [4]. Therefore, alternative concepts for magnetism control, that allow one to use low electric fields at elevated temperatures, are actively pursued. Moreover, an additional requirement for applications is the integration of the magnetic system into an electronic device compatible with current semiconductor technology.

Hybrid systems that combine thin ferromagnetic (FM) films with semiconducting (SC) layers are promising for unifying magnetism and electronics, which may allow all-in-one-chip solutions for computing. To that end, the hybrids need to show magnetic order as a ferromagnet, while remaining electrically reconfigurable as a semiconductor [6, 7]. By now, ferromagnetic proximity effects were revealed optically [8, 9] and electrically [10]. Further, electrical measurements using the anomalous Hall effect demonstrated that the *p-d* exchange interaction of the magnetic atoms in a ferromagnetic film (the *d*-system) with a two-dimensional hole gas (2DHG, the *p*-system) in a semiconductor quantum well (QW) induces an equilibrium spin polarization of the QW holes [10]. Optical studies [8, 9] showed polarized photoluminescence (PL) from the QW located a few nanometers apart from the FM. However, care has to be exercised in the interpretation of the FM proximity effect, when electrons and holes are present in non-equilibrium: a previous study [11] had demonstrated that under optical excitation an alternative mechanism exists involving spin-dependent capture of charge carriers from the SC into the FM, representing a dynamical spin polarization effect in contrast to the exchange-induced equilibrium polarization. All these mechanisms are based on wavefunction overlap and, therefore, lead to short-range proximity effects.

A novel type of long-range FM proximity effect was reported recently for a hybrid Co/CdTe structure [12]. It is manifested by the spin polarization of holes bound to shallow acceptors in a nonmagnetic CdTe quantum well due to an effective long-range *p-d* exchange interaction that is not related to the penetration of the electron wavefunction into the FM layer. This interaction was conjectured to be mediated by elliptically polarized phonons with energy close to the magnon-phonon resonance in the FM. The long-range exchange constant was directly measured by spin-flip Raman scattering (SFRS) in Ref. [13]. However, no electric control of this exchange coupling has been demonstrated so far.

Here, we show that application of an electric field across the structure changes the strength of the long-range *p-d* exchange coupling between the FM and the SC, namely the holes bound to acceptors in the quantum well. The coupling is controlled by the band bending in the quantum well region, becoming most efficient in the case of flat bands. The effective magnetic field of the exchange interaction reaches 2.5 T and can be turned on and off by application of ~ 1 V bias across the heterostructure. The control is not related to a spatial redistribution of wavefunctions and, therefore, cannot be explained using the standard model of exchange interaction. In contrast, it can be well described in the framework of the exchange mechanism mediated by elliptically polarized phonons. The applied voltage varies the heavy-light hole transition to which the phonons couple, bringing it in and out of resonance with the magnon-phonon resonance of the FM. Doing so, the effective exchange coupling strength in the hybrid system is tuned electrically without any power consumption, using field strengths of about $10^4$ V/cm only, which is a few orders of magnitude reduced in comparison to non-semiconductor systems. Our results pave the way for integration of electrically tunable magnetism into semiconductor electronics.

**1. Ferromagnetic proximity effect in steady state**

The investigated Co/(Cd,Mg)Te/CdTe/(Cd,Mg)Te/CdTe:I/GaAs hybrid structure was grown by molecular-beam epitaxy on a GaAs substrate followed by a conducting CdTe:I layer (10 μm thickness, iodine-doped with donor concentration of ~$10^{18}$ cm$^{-3}$) as sketched in the layer-by-layer design in Fig. 1a. The QW is formed by a 10 nm CdTe layer sandwiched between layers of 0.5 μm CdMgTe and 8 nm CdMgTe (the spacer). On top of this structure, the 4 nm thick cobalt film was deposited. A mesa of 5 mm diameter was lithographically patterned by deep etching, so that an applied voltage drops between the Co and CdTe:I layers. Figure 1b schematically shows the band diagram of the structure. The current-voltage characteristics $I(U)$ in Fig. 1c reflect the typical behavior of a Schottky diode shifted downwards along the *I*-axis due to the photovoltaic effect.

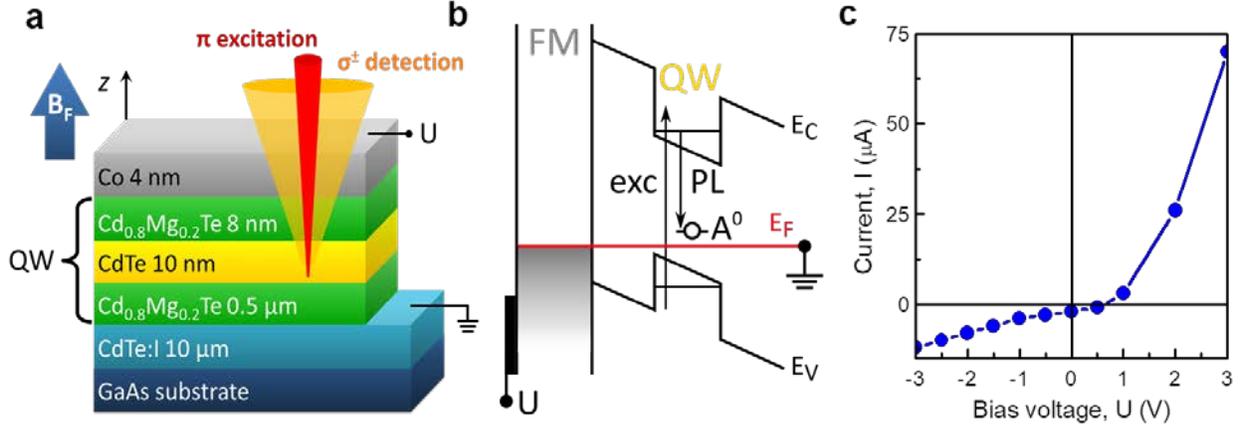

Figure 1. Sample characterization. (a) Schematic device architecture and geometry of the cw experiment. (b) Schematic presentation of the band diagram at $U = 0$, where $E_V$ and $E_C$ denote valence and conduction bands, respectively, and $E_F$ indicates the Fermi level. (c) Current-voltage characteristics $I(U)$ under excitation with laser light in cw mode.

We study the ferromagnetic proximity effect using polarized photoluminescence (PL) spectroscopy in the continuous wave (cw) mode. The sample is excited by linearly polarized (π) light and the degree of circular polarization $\rho_c^\pi$ of the photoluminescence from the quantum well is detected. The value of $\rho_c^\pi$ does not depend on the orientation of the linear laser polarization. To magnetize the interfacial ferromagnet, which is responsible for the FM proximity effect [12], we apply a magnetic field $B_F$ in the Faraday geometry parallel to the growth axis of the heterostructure (z-axis, Fig.1a).

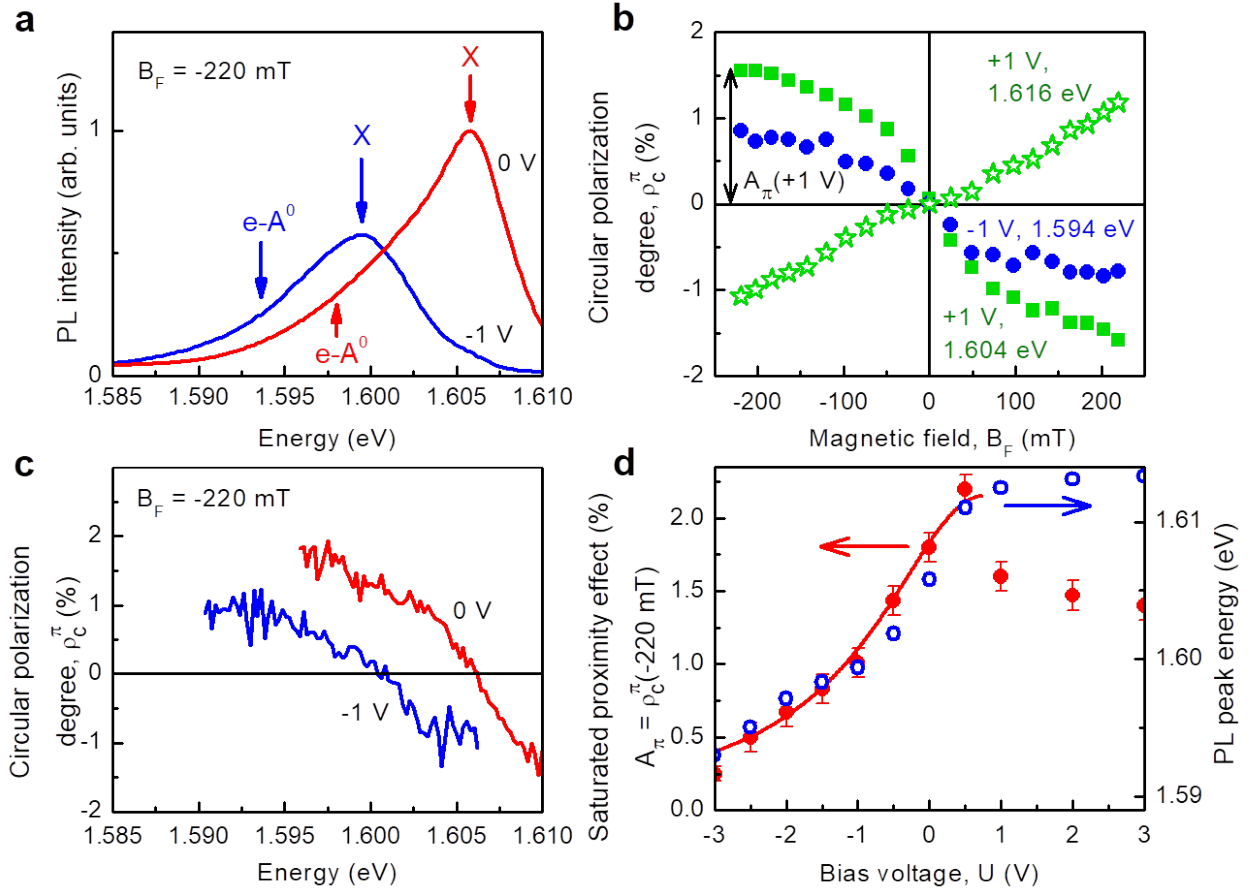

Figure 2. Polarization spectroscopy in the cw regime at different bias voltages. (a) PL spectra at $U = 0$ (red) and $-1$ V (blue). (b) Magnetic field dependences of the circular polarization degree $\rho_c^\pi(B_F)$ measured at $U = +1$ V, $\hbar\omega_{PL} = 1.604$ eV (green squares), $U = -1$ V, $\hbar\omega_{PL} = 1.594$ eV (blue circles) and $U = +1$ V, $\hbar\omega_{PL} = 1.616$ eV (green stars), using linearly polarized excitation. The arrow indicates the amplitude $A_\pi(+1V)$ of the FM proximity effect for $U = +1$ V at 1.604 eV detection energy. (c) Spectral dependences of $\rho_c^\pi$ for $U = 0$ V and $-1$ V. (d) Amplitude $A_\pi(U)$ of the FM proximity effect (red circles) and the energy of the PL peak $\hbar\omega_{max}(U)$ (blue circles) as functions of the gate voltage. The red curve is a fit using Eq. (4.4) with $A = 2.2$ %, $U_0 = 0.8$ V and $U_1 = 1.8$ V. Excitation was done with photon energy 1.691 eV (below the band gap of the CdMgTe barriers) using a power density of 4 W/cm$^2$. In all cases the magnetic field was $B_F = -220$ mT (except of (b)) at temperature $T = 2$ K.

The red curves in Figs. 2a,c correspond to zero bias, the blue ones to $U = -1$ V, all taken at $B_F = -220$ mT. In the PL spectrum (Fig. 2a), two features are observed. The PL band at higher photon energies (X) corresponds to the recombination of the exciton in the QW, while the low-energy tail (e-A$^0$) is attributed to the recombination of an electron with a hole bound to a shallow acceptor in the QW [12]. At zero bias $U = 0$ (red curves) the photoluminescence reveals a polarization of about 2 % at the acceptor band (Fig. 2c), undergoing a sign change towards the exciton emission. Application of reverse bias (blue curves at $U = -1$ V) shifts the entire spectrum to lower energies by 7 meV due to the strong bending of the energy bands by the (static) Stark effect. Simultaneously the PL intensity decreases due to the separation of electron and hole in the QW by the electric field, leading to a reduced transition matrix element. Here the polarization degree around the acceptor emission is reduced to about 1% (Fig. 2c).

Next, the magnetic field dependencies of $\rho_c^\pi(B_F)$ were measured in the Faraday geometry for different fixed biases $U$. The FM proximity effect is assessed by the degree of circular polarization of the e-A$^0$ PL versus magnetic field $B_F$. The degree $\rho_c^\pi(B_F)$ saturates in the field range of 150-200 mT (green squares for $U = +1$ V and blue circles for $U = -1$ V). Since the spectral positions of the emission bands are sensitive to the applied voltage $U$, the polarization was detected at the photon energy $\hbar\omega_{PL}$ corresponding to the maximum polarization of the e-A$^0$ transition (Fig. 2c). The magnitude of the FM proximity effect is given by the saturation polarization at the acceptor band $A_\pi \equiv \rho_c^\pi(B_F = -220$ mT$)$ which is larger for +1 V than for the reverse bias of −1 V. In contrast to the e-A$^0$ transition, the polarization $\rho_c^\pi(B_F)$ near the exciton

PL maximum depends linearly on magnetic field across the whole scanned field range without any saturation (green stars in Fig. 2b, $U = +1$ V, $\hbar\omega_{PL} = 1.616$ eV), independent of the applied bias $U$. The linear dependence of $\rho_c^\pi(B_F)$ originates from the X splitting in two lines with opposite circular polarization due to the Zeeman effect [12]. The linear $B_F$-dependence of $\rho_c^\pi(B_F)$ indicates that the proximity effect is absent for the valence band holes that contribute to the exciton within its lifetime [12].

Figure 2d shows the dependence of the saturation polarization $A_\pi(U)$ (red circles) and the energy of the PL peak $\hbar\omega_{max}(U)$ (blue circles) as function of applied bias. The energy $\hbar\omega_{max}(U)$ increases with bias due to the reduction of the inclination of bands and consequently of the static Stark effect. The external positive bias decreases the built-in electric field. It is canceled at $U > +0.5$ V (flat band conditions) as evidenced also by a steep increase of current through the device (Fig. 1c). In this case the voltage drop is redistributed all over the structure plane (Fig. 1a), so that a voltage increase does not lead to a further band inclination. For $U \leq +0.5$ V the bias increases the built-in electric field and we observe a striking correlation between the voltage dependences of the magnitude of the FM proximity effect $A_\pi(U)$ and the peak position $\hbar\omega_{max}(U)$ (see Fig. 2d). In turn, for $U > +0.5$ V the FM proximity effect decreases about 1.5 times reaching the level of 1.5%. This drop is attributed to the appearance of additional holes in the valence band of the QW that have negligible $p$-$d$ exchange coupling (for details see supplementary section S2).

Here, we concentrate on the origin of the voltage dependence $A_\pi(U)$ for $U < +0.5$ V. The FM proximity effect originates from the effective $p$-$d$ exchange interaction $\frac{1}{3}\Delta_{pd}J_z$ between the interfacial FM [12] and the acceptor-bound QW holes with the momentum projections $J_z = \pm 3/2$ onto the $z$-axis. The saturation amplitude $A_\pi$ of the PL polarization is caused by the spin polarization $P_A$ of $A^0$ when the FM is completely magnetized

$$A_\pi = P_A = -\frac{\tau_A}{\tau_A + \tau_{sA}} \frac{\Delta_{pd}}{2k_B T} \ . \qquad (1.1)$$

Here $\tau_A$ is the lifetime and $\tau_{sA}$ is the spin relaxation time of the heavy holes on acceptors, $k_B$ is the Boltzmann constant, $T$ is the lattice temperature, and $\Delta_{pd}$ is the spin splitting of the $\pm 3/2$ levels in the effective magnetic field of the $p$-$d$ exchange interaction. A positive sign of $\Delta_{pd}$ implies that the $-3/2$ state is energetically favorable [12]. The polarization $P_A$ can depend on the bias $U$ through four dependencies: 1) the ratio of the times $\tau_A/\tau_{sA} = f(U)$, 2) the constant $\Delta_{pd}(U)$ of the $p$-$d$ exchange, 3) the lattice heating $T(U)$ by electrical current, and 4) the injection of spin-polarized carriers from the FM. Heating can be excluded because the electrical power in our experiment was two orders of magnitude (<40 μW) smaller than the optical power. Heating due to the injection of hot holes and the spin injection option can be ruled out because the amplitude $A_\pi(U)$ does not follow the electric current $I(U)$ (Fig.1c). Time-resolved PL and spin flip Raman scattering experiments demonstrate that the $\Delta_{pd}(U)$ dependence is the main origin of $A_\pi(U)$.

## 2. Electrical control of the kinetics of ferromagnetic proximity effect

Time-resolved PL enables one to measure the kinetics of the PL intensity and thereby the emergence of spin polarization induced by the magnetized FM layer (ferromagnetic proximity effect) after optical excitation of non-polarized charge carriers with linearly polarized laser pulses (Fig. 3a). In Ref. [12] we demonstrated that the exciton PL does not reveal the FM proximity effect. The PL intensity decays much faster (a few 100 ps) than the rise of the e-$A^0$ PL (~ 2 ns). Here, the same scenario is realized. Due to the dominant contribution of the exciton to the total PL signal especially during the first few hundred picoseconds (black curve in Fig. 3a), it is necessary to wait for 500-700 ps, until the excitons have mostly recombined, to reliably evaluate the FM proximity effect. Figure 3a (orange circles) shows the evolution of the circular polarization starting from time delays of about 700 ps. The blue dashed curve in Fig. 3a is the fit of the data according to $\rho_c^\pi(t) = \rho_{sA}[1 - \exp(-t/\tau_{sA})]$ with the amplitude $\rho_{sA} = 13\%$, and rise time $\tau_{sA} = 1.3$ ns. This measurement was carried out at $U = 0$. From kinetic measurements of the FM proximity effect at other bias voltages one can determine the dependences $\rho_{sA}(U)$ and $\tau_{sA}(U)$.

Because of the delay waiting for exciton recombination, the dependence $\rho_{sA}(U)$ can be measured more accurately than $\tau_{sA}(U)$. $\rho_{sA}(U)$ has a straightforward interpretation from which $\Delta_{pd}$ can be inferred. Equilibrium occurs at delay times much longer than the spin relaxation time $\tau_{sA}$. Then the recombining electrons are characterized by the equilibrium spin state given by the Boltzmann distribution. Therefore, the polarization amplitude

$$\rho_{sA}(U) = -\frac{\Delta_{pd}(U)}{2k_B T} \qquad (2.1)$$

is determined solely by the ratio of the exchange constant to the thermal energy and does not depend on the ratio of lifetime and relaxation time. Thus, the dependence $\rho_{sA}(U)$ is determined exclusively by $\Delta_{pd}(U)$. Figure 3b shows that $\rho_{sA}(U)$ has a peak near $U = +1$ V, similar to the cw data (compare with Fig. 2d). Therefore, the time-resolved PL demonstrates that the exchange constant between the magnetic ions and the holes bound to acceptors (rather than the $\tau_A/\tau_{sA}$ ratio) is controlled by the electric field.

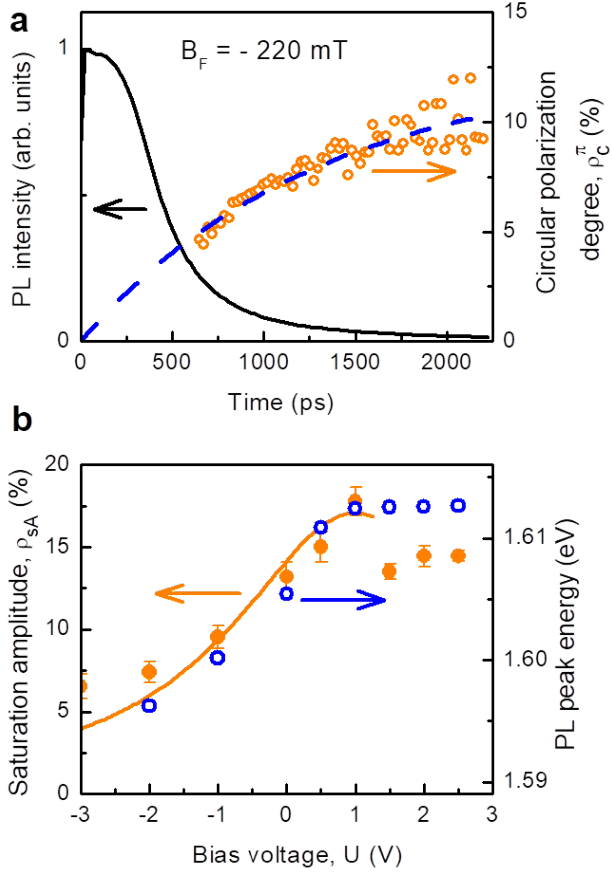

Figure 3. (a) The kinetics of the PL intensity (black solid line) and the degree of circular polarization of the PL (open orange circles) integrated in the spectral region of e-$A^0$ at $U = 0$. Blue dashed line is a fit to the data. (b) Dependence of $\rho_{sA}(U)$ of the FM proximity effect (full orange circles) and the position of the PL peak $\hbar\omega_{max}(U)$ (open blue circles) on the bias voltage $U$. The solid curve is calculated using Eq. (4.4) with $A = 17\,\%$, $U_0 = 1.0$ V and $U_1 = 2$ V.

## 3. Determination of the *p-d* exchange constant by spin-flip Raman scattering

SFRS under resonant excitation of the exciton bound to an acceptor ($A^0X$) complex can be used as a reliable tool to determine directly the magnitude of the FM induced exchange splitting [14]. In tilted magnetic field three spin flip processes occur as discussed in the supplementary section S3. Using excitation with $\sigma^-$ polarization and detection with $\sigma^+$ cross-polarization, each of these processes results in a Stokes shift of the Raman signal by a characteristic energy. The first process is associated with a double spin flip of electron and hole with participation of an acoustic phonon. In presence of the *p-d* exchange interaction the Stokes shift is given by

$$\Delta_S^{DSF} = \hbar\omega_1 - \hbar\omega_2 = \mu_B(|g_e| + |g_A|)B - \Delta_{pd}, \tag{3.1}$$

where $\hbar\omega_1$ and $\hbar\omega_2$ are the energies of the incoming and scattered photons, $\mu_B$ is the Bohr magneton, $g_e$ and $g_A$ are the g-factors of the electron and the acceptor-bound hole, respectively. Another SFRS-contribution is represented by the single spin flip of the acceptor bound hole. In the presence of the *p-d* exchange interaction, the corresponding Stokes shift is

$$\Delta_S^{SSF} = \hbar\omega_1 - \hbar\omega_2 = \mu_B |g_A| B - \Delta_{pd}. \tag{3.2}$$

The third process is associated with the spin flip of the electron in the excited $A^0X$ complex resulting in a Stokes shift determined solely by the Zeeman splitting, $\mu_B |g_e| B$.

The Raman spectrum (Fig. 4a) under resonant excitation of the $A^0X$ complex (1.600 eV) shows the broad SFRS line "h", which is associated with the single hole spin flip process. The "e+h" line corresponds to the double spin flip process. Finally, there is also the line "e", which corresponds to the electron spin flip. The sum of the energy shifts of the "e" and "h" peaks gives the energy of "e+h" peak. The energies of all three SFRS lines change linearly with applied magnetic field (see Fig. 4b). However, when the magnetic field is extrapolated to zero, the Stokes shift of the line "e" tends to zero, while the lines "h" and "e+h" show a negative offset. This means that both SFRS lines "h" and "e+h" are influenced by the exchange interaction with the FM layer and, thus, can be used to assess the effect of gate voltage on the exchange coupling strength. The zero field offset represents a direct measurement of the exchange constant $\Delta_{pd}$. The dependence of $\Delta_{pd}$ on the applied voltage $U$ is shown in Fig. 4c by the green circles. It correlates well with the voltage dependence of the PL peak position for $U < + 0.5$ V. The maximum value of $\Delta_{pd} \approx 150$ µeV occurs in the flat band regime followed by a fast drop with increasing $U$ to a value of 50 µeV, where it remains constant for $U > +0.5$ V, similar to Figs. 2d and 3b. The splitting value $\Delta_{pd} \approx 150$ µeV corresponds to an effective magnetic field of exchange interaction of 2.5 T (the Lande g-factor $|g_A| \approx 1$). Thus, the exchange interaction can be turned on and off by the application of ~ 1 V bias across the heterostructure of ≤ 1 μm thick, i.e. by an electric field of about $10^4$ V/cm, which is a few orders of magnitude less than previously reported in other systems [4].

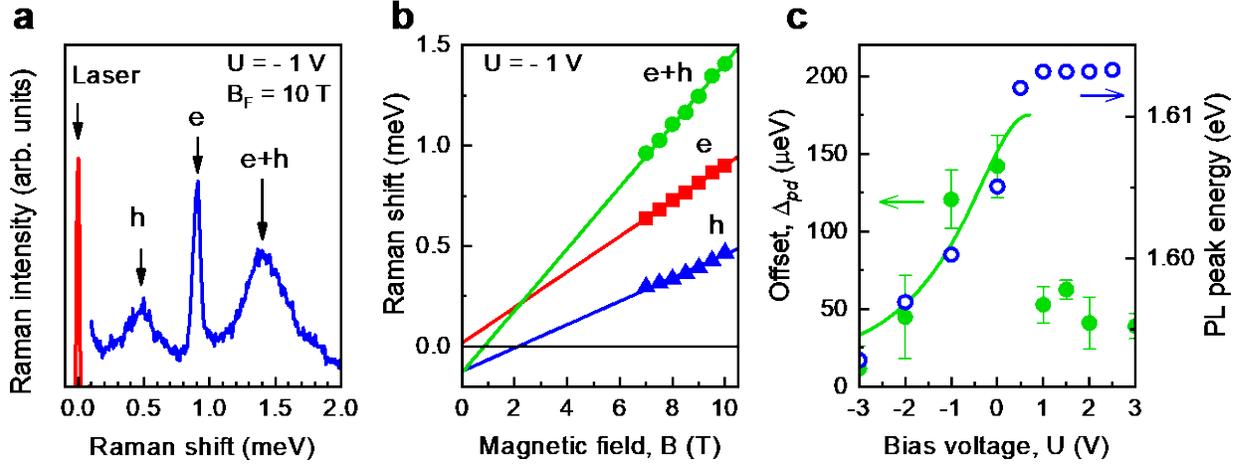

Figure 4. (a) Spin flip Raman signal under resonant excitation of the $A^0X$ complex at a magnetic field $B_F = 10$ T tilted by 20° with respect to the sample growth direction. Line "h" corresponds to a single acceptor hole spin flip process, "e+h" is a double spin flip process, and "e" is a spin flip of a conduction band electron. (b) Dependence of the Stokes shifts of the three lines on the magnetic field. (c) Dependence of the exchange constant $\Delta_{pd}$ on the applied voltage $U$ for the hole bound to an acceptor ("h", green solid circles) and the corresponding dependence of the energy of the PL peak $\hbar\omega_{max}(U)$ (blue open circles). The solid curve is calculated from Eq. (4.4) with $A = 175$ μeV, $U_0 = 0.7$ V and $U_1 = 1.8$ V.

## 4. Model

The main finding of this work is the electric field control of the long-range exchange interaction in a hybrid ferromagnet-semiconductor structure. It can be well explained in the frame of the indirect *p-d* exchange mechanism mediated by elliptically polarized phonons, which represents *the phononic ac Stark effect* [12]. Elliptically polarized phonons exist in ferromagnets near the energy $E_{mp}$ of the magnon-phonon resonance [15]. The FM proximity effect is based on the spin-orbit interaction of the spin of acceptor-bound holes in the QW with the nonzero angular momentum of these acoustic phonons as shown in the energy diagram of Fig. 5 in electron-electron representation.

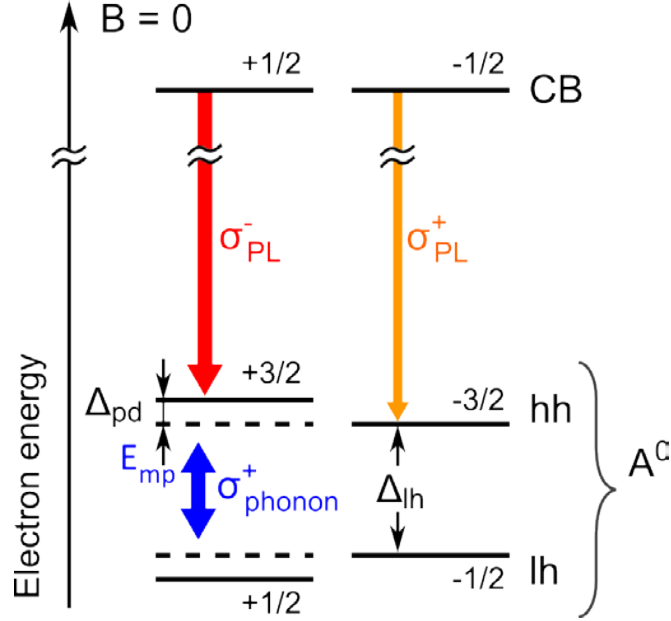

Figure 5. Energy scheme of the spin states of a conduction band electron (CB) and a neutral acceptor ($A^0$). The acceptor states with spin projections +3/2 and +1/2 are shifted with respect to the unperturbed energy states due to the phononic ac Stark effect mediated through the elliptically polarized phonons emitted by the magnetized FM with preferential $\sigma^+$ polarization. The unperturbed +3/2 and +1/2 energy states are indicated by dashed lines.

The neutral acceptor states are split in two doublets with angular momentum projections along the *z*-axis equal to ±3/2 (heavy hole states) and ±1/2 (light hole states). In this quadruplet in its ground state, the ±1/2 acceptor levels are populated with electrons, while the ±3/2 states are mostly empty. Interaction with circularly polarized phonons leads to an energy shift of the acceptor levels and a lifting of the doublets' degeneracy associated with the angular momentum projection. The effect is maximal near the magnon phonon resonance, where the energy $E_{mp} \approx 1$ meV [15] is close to the energy splitting $\Delta_{lh} \approx 1$ meV [14] between the heavy- and light-hole acceptor states. The experimental results of Ref. [12] demonstrate that a magnetization of the FM layer along the *z*-axis leads to negative circular polarization of the e-A$^0$ optical transition, i.e. the state with +3/2 projection is shifted to higher energies with respect to the -3/2 level (see Fig. 5). Therefore, the unpolarized electrons from the conduction band mainly recombine into the empty +3/2 electronic states and the resulting emitted photons are $\sigma^-$ polarized. Such a level sequence is obtained when the elliptically polarized phonons have preferential $\sigma^+$ component and $\Delta_{lh} > E_{mp}$. Indeed, in this case the coupling is established between the electronic ground state +1/2 of the acceptor in the presence of *N* phonons and the excited state +3/2 with *N*−1 phonons. In case of $\Delta_{lh} > E_{mp}$ the energy levels repel each other. This is extracted from the experimental results of this work and the conclusions of Ref. [16], where application of a static electric field *E* increases $\Delta_{lh}$, leading to an increase of the detuning $\Delta_{lh}(E) - E_{mp}$, resulting in a decrease of the interaction between the QW and the FM layer.

Similarly to the optical ac Stark effect [17], $\Delta_{pd}$ can be calculated in second order perturbation theory

$$\Delta_{pd}(E) = E_{+3/2} - E_{-3/2} = \frac{|H_{ph-h}|^2}{\Delta_{\ell h}(E) - E_{mp}} P_{phon}^c . \tag{4.1}$$

Here $H_{ph-h}$ is the matrix element of the interaction of the acceptor spin with the transverse acoustic phonons. Similarly to the circular polarization degree $\rho_c$ of light the degree of phonon circular polarization is $P_{phon}^c = \frac{N_+ - N_-}{N_+ + N_-}$, with the number $N_+$ ($N_-$) of right (left) circularly polarized phonons (with energy close to the magnon-phonon resonance energy). Analogously to the optical ac Stark effect the constant $\Delta_{pd}$ is determined by the detuning $\Delta_{lh}(E) - E_{mp}$ of the phonon energy $E_{mp}$ (analog of the photon energy $\hbar\omega$) from the splitting $\Delta_{lh}(E)$ (analog of the energy of the optical transition in an atom). The sign of $P^c_{phon}$ in the vicinity of the magnon-phonon resonance depends on the sign of the projection of the magnetization component of the interfacial ferromagnet onto the $z$-axis. The electric field $E$ across the QW increases $\Delta_{\ell h}(E)$ due to the static quadratic Stark effect. For small values of the electric field directed along the $z \parallel [001]$ axis

$$\Delta_{\ell h}(E) = \Delta_{\ell h}(0) + a_8 E^2 . \tag{4.2}$$

The first term $\Delta_{\ell h}(0)$ on the right hand side in Eq.(4.2) corresponds to the splitting in zero electric field due to quantum confinement, while the second term is the correction due to the Stark effect. The parameter $a_8 > 0$ determines the static susceptibility of a neutral acceptor and is known only for shallow acceptors in Si [16]: $a_8 \sim 10^{-10}$ eV·cm$^2$/V$^2$. Then an electric field strength as low as $10^3$ V/cm induces an energy shift of $\sim 0.1$ meV, which is comparable to the initial detuning $\Delta_{lh}(0) - E_{mp}$. Hence, a relatively small electric field can control the exchange coupling constant. Substituting Eq. (4.2) into Eq.(4.1), we obtain

$$\Delta_{pd}(E) = \frac{|H_{ph-h}|^2}{\Delta_{\ell h}(0) - E_{mp} + a_8 E^2} P^c_{phon} . \tag{4.3}$$

Since the static susceptibility $a_8 > 0$ [16], our results demonstrate that the $\Delta_{pd}$ value is maximum for the case of flat bands ($E=0$), where we have $\Delta_{lh}(0) - E_{mp} > 0$. The experiment [12] shows that $\Delta_{pd} > 0$ for $B_F > 0$, and therefore, as mentioned before, $P^c_{phon} > 0$, i.e. the phonons are mainly $\sigma^+$ polarized in agreement with Fig. 5.

We fit the data assuming that in a small range of reverse bias the electric field is proportional to the applied voltage $E \propto (U - U_0)$, where $U = U_0$ corresponds to the flat band conditions, and drops entirely within the undoped region of ≤ 1 μm thickness (Fig. 1a). Thus from Eq. (4.3) we get a Lorentz curve

$$f(U) = \frac{A}{1 + (U - U_0)^2 / U_1^2} \quad (4.4)$$

with halfwidth $U_1$. The amplitude $A$ gives the magnitude of the effect under flat bands conditions, and has different dimensions for different experimental techniques. For SFRS the amplitude $A$ in Eq. (4.4) gives the value of $\Delta_{pd}(U)$ in meV. The solid curve in Fig. 4c fits the results of the SFRS measurements well with $A = 175$ μeV, $U_0 = 0.7$ V and $U_1 = 1.8$ V. For the polarization measurements the amplitude $A$ in Eq. (4.4) is dimensionless. The solid line in Fig. 2d fits the polarization amplitude data in cw mode for $A = 2.2$ %, $U_0 = 0.8$ V and $U_1 = 1.8$ V. The time-resolved PL data of the polarization kinetics in Fig. 3b are described by $A = 17$ %, $U_0 = 1.0$ V and $U_1 = 2$ V. These fit parameters demonstrate good agreement. Therefore, the results of all three experimental techniques are explained within the model of the phononic ac Stark effect. Our results demonstrate that $U_1 \approx 1.5$ V is enough to switch the *p-d* interaction off, i.e. we obtain indeed a low-voltage control of magnetism.

**Concluding remarks**

At first glance the low-voltage control of the long-range exchange coupling looks surprising. Indeed, common sense suggests that application of an electric field in a direction that attracts holes inside the QW towards the FM should enhance the *p-d* exchange interaction. In contrast to that, the *p-d* exchange coupling decreases when reverse bias voltage (negative potential at the top electrode) is applied. This finding supports the earlier conclusions on the origin of the long-range *p-d* exchange interaction which is not related to the penetration of the electronic wavefunctions into the FM layer [12]. Our data rather demonstrate and confirm the new suggested mechanism of the phononic ac Stark effect for electric field control of the exchange interaction, which is essentially different from traditional concepts. The coupling strength $\Delta_{pd}(E)$ correlates with the band bending in the quantum well region and can be explained in the frame of the exchange coupling mediated by the elliptically polarized phonons – the phononic ac Stark effect. The electric field changes the detuning of the heavy-light hole energy splitting of the QW acceptor with respect to the magnon-phonon resonance energy in the FM. This result corroborates the feasibility of electrical control of the exchange interaction in hybrid ferromagnet-semiconductor nanostructures and can be potentially used for applications such as electric field effect magnetic memories. From a fundamental point of view, our achievement opens a principally new way for the control of magnetic interactions via the gate tunable phononic ac Stark effect that can be extended to various magnetic systems.

**Methods: Experiment and sample**

The sample of Co/(Cd,Mg)Te/CdTe/(Cd,Mg)Te/CdTe:I/GaAs (Fig. 1a) was grown on a (100)-oriented GaAs substrate by molecular-beam epitaxy. The buffer between the substrate and the quantum well is a 10 μm layer of conductive CdTe doped with iodine (donor concentration of the order of $10^{18}$ cm$^{-3}$). The quantum well consists of a 0.5 μm wide (Cd,Mg)Te barrier layer, a 10 nm CdTe layer and an 8 nm CdMgTe spacer. On top a 4 nm thick cobalt layer is deposited. In order to make electrical contact to the CdTe:I buffer layer a 5 mm in diameter mesa was etched into the structure to a depth of more than 0.8 μm. One contact is wired to the CdTe:I buffer layer, and the second contact is located at the cobalt surface.

For continuous wave polarization-resolved photoluminescence spectroscopy the sample was excited by the linearly polarized (π) light of a titanium-sapphire laser. In order to avoid sample heating the laser power was kept below 4 mW/cm². The degree of circular polarization $\rho_c^\pi = (I_+^\pi - I_-^\pi)/(I_+^\pi + I_-^\pi)$ of the PL from the QW was detected, where $I_{\sigma+}^\pi$ and $I_{\sigma-}^\pi$ are the intensities of the σ⁺ and σ⁻ components with right and left circular polarization, respectively. The polarization degree $\rho_c^\pi$ does not depend on the orientation of the laser polarization. To magnetize the interfacial ferromagnet, a small magnetic field $B_F$ (of the order of 100 mT) was applied in the Faraday geometry normal to the structure plane using a resistive magnet. The measurements were carried out at a temperature of 2 K.

Time-resolved PL allows one to obtain information about the transient processes of decay of the photoexcited carriers and their spin relaxation. Here, the sample is excited by short optical pulses with a central photon energy of 1.69 eV using a self-mode-locked titanium-sapphire laser with a repetition frequency of 75.75 MHz. The pulse duration was 150 fs, the spectral width of the laser was 10 nm, and the average pump density was ~ 4 W/cm². The PL was dispersed with a 0.5 m focal length single monochromator to which a streak camera was attached for detection. The overall time resolution was about 20 ps. The experiments were carried out at a temperature of 2 K.

The coherent spin dynamics was measured by conventional time-resolved pump-probe Kerr rotation using a titanium-sapphire laser generating 1.5 ps pulses at the repetition frequency of 75.6 MHz (repetition period $T_R = 13.2$ ns). Electron spin coherence was generated along the growth $z$-axis of the sample by circularly polarized pump pulses. The polarization of the beam was modulated between $\sigma^+$ to $\sigma^-$ by a photo-elastic modulator operated at a frequency of 50 kHz. In order to avoid electron heating and delocalization effects the average pump density was kept at low levels $\leq 5$ W/cm$^2$. The probe beam was linearly polarized. The angle of its polarization rotation ($\theta_K$) or the ellipticity after reflection of the beam from the sample was measured by a polarization sensitive beamsplitter in conjunction with a balanced photodetector. Pump and probe beams had the same photon energy and were tuned to the energy of the exciton resonance. The sample was placed in the temperature insert of a vector magnet cryostat containing three superconducting split coils oriented orthogonally to each other. This magnet allows one to ramp the magnetic field up to 3 T and to carry out measurements with different orientations of the magnetic field relative to the sample at temperatures from $T = 1.7$ K up to 300 K.

The spin-flip Raman scattering (SFRS) experiments were performed using resonant excitation with a cw laser at photon energies corresponding to the PL band of excitons bound to neutral acceptors (A$^0$X) which is located about 1 meV below the exciton optical transition (see Fig. 2a). We used an oblique backscattering Faraday geometry where the excitation/detection beams and the magnetic field were parallel to each other, while the sample growth $z$-axis was tilted by 20 degrees with respect to the magnetic field direction. The Raman shift was measured at a temperature of 2 K in magnetic fields up to 10 T in crossed circular polarizations for excitation and detection. The SFRS spectra were dispersed by a Jobin-Yvon U-1000 monochromator equipped with a cooled GaAs photomultiplier.

**Acknowledgments:**

We acknowledge support by the Deutsche Forschungsgemeinschaft and Russian Foundation for Basic Research in the frame of the ICRC TRR 160 (Project C2, B2 and A1). The partial financial support from the Russian Foundation for Basic Research Grant No. 15-52-12017 NNIOa and 19-52-12034 NNIOa is acknowledged. The research in Poland was partially supported by the National Science Centre (Poland) through grants Harmonia and UMO 2017/25/B/ST3/ 02966 and by the Foundation of Polish Science through the IRA Programme co-financed by EU within SGOP.

**Additional information**

The authors declare no competing financial interests.

**Individual Contributions**

V.L.K., I.V.K., I.A.A., V.F.S., E.A.Z., E.K., O.S.K., D.K. performed the experiments and analyzed the data. V.L.K. developed the theoretical model. G.K., M.W., and T.W. fabricated the samples. N.D.I., N.M.L. patterned the mesa structure, and T.K. prepared thed contacts. V.L.K., I.A.A., D.R.Y., Yu.G.K, and M.B. co-wrote the paper. All authors discussed the results and commented on the manuscript.

Supplementary material

# Voltage control of long-range *p-d* exchange coupling in a ferromagnet-semiconductor quantum well hybrid structure

V.L. Korenev[1,*], I.V. Kalitukha[1], I.A. Akimov[1,2,*], V.F. Sapega[1], E.A. Zhukov[1,2], E. Kirstein[2], O.S. Ken[1], D. Kudlacik[2], G. Karczewski[3], M. Wiater[4], T. Wojtowicz[4], N.D. Ilyinskaya[1], N.M. Lebedeva[1], T.A. Komissarova[1], Yu.G. Kusrayev[1], D.R. Yakovlev[1,2], and M. Bayer[1,2]

[1]Ioffe Institute, Russian Academy of Sciences, 194021 St. Petersburg, Russia

[2]Experimentelle Physik 2, Technische Universität Dortmund, D-44227 Dortmund, Germany

[3]Institute of Physics, Polish Academy of Sciences, PL-02668 Warsaw, Poland

[4]International Research Centre MagTop, Institute of Physics, Polish Academy of Sciences

PL-02668 Warsaw, Poland

## S1. Pump-probe measurements in pulsed mode

Near the QW exciton resonance of the studied hybrid structure, pump-probe experiments performed in magnetic field are used to measure the frequency of the Larmor precession of conduction band electrons $\omega_e(B)$ and valence band holes $\omega_h(B)$ [13]. Under excitation with $\sigma^+$ polarized pump pulses, electrons and holes spin-polarized along the growth axis of the structure appear. Their spins precess about a perpendicular magnetic field as manifested by an oscillatory signal being proportional to the $z$-component of the spins. Electrons and valence band holes precess with different frequencies due to the strong difference between their Landé $g$-factors. The electron $g$-factor in the CdTe QW is practically isotropic, while that of the valence band hole is strongly anisotropic [S1]. Here, the experiment was performed in a tilted magnetic field, so that the field component along the sample growth direction ($z$-axis) induces the magnetization of the interfacial FM, while its transverse component causes spin precession of the charge carriers in the QW. The details are described in Ref. [13]. We monitor the precession of the valence band hole spins induced by the magnetic field for $U = +1$ V. The measurements were carried out in the magnetic fields of 0.5–3 T tilted by $\theta = 80°$ with respect to the $z$-axis, so that its $z$-component was in the range of 100–520 mT. This is sufficient to magnetize the interfacial FM along the easy $z$-axis and to cause the FM proximity effect [12]. An example of spin oscillations in a magnetic field of $B_V = 1$ T in the Voigt geometry is shown in Fig. S1a. A superposition of electron (fast) and hole (slow) oscillations is observed. Fitting with two exponentially damped harmonic functions allows one to determine the oscillation frequencies. The magnetic field dependences of the energy of the hole Zeeman splitting $\hbar\omega_h(B)$ for $\theta = 80°$ (red circles) and 90° (blue circles) are plotted in Fig. S1b.

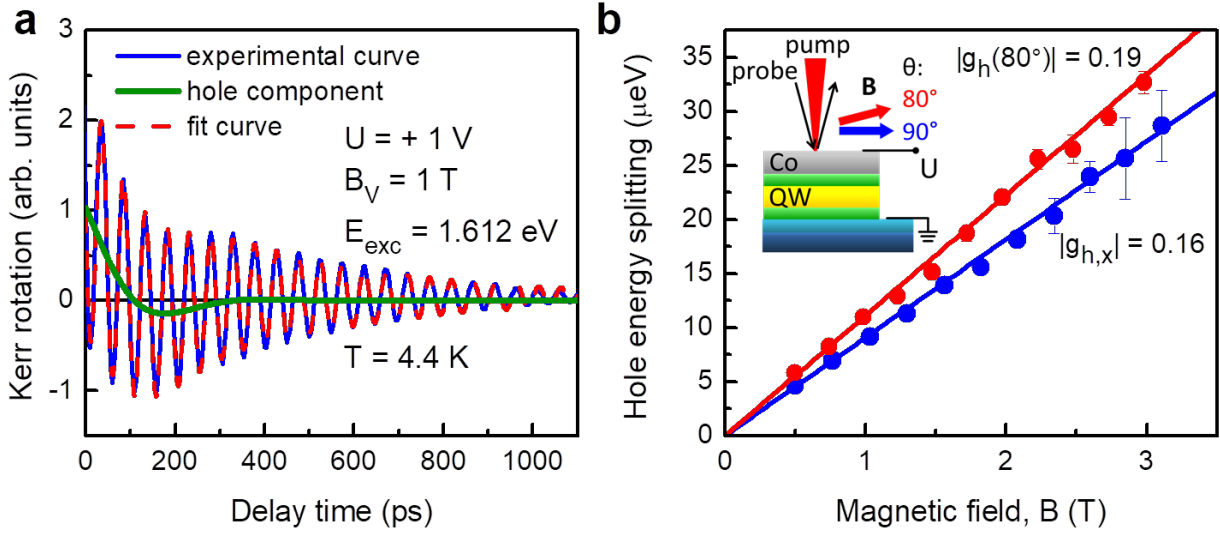

Figure S1. (a) Oscillations of the Kerr-rotation angle in a magnetic field of $B_V = 1$ T in the Voigt geometry for a forward bias $U = +1$ V. (b) Dependence of the energy of the Zeeman splitting of the hole on the magnetic field for the Voigt geometry ($\theta = 90°$, blue circles) and for a tilted field ($\theta = 80°$, red circles). The voltage $U = +1$ V, the photon energy of the pump and the probe is 1.612 eV, the power density of both the pump and the probe is 5 W/cm². $T = 4.4$ K.

Figure S1b shows that the g-factor of the hole is anisotropic: $|g_{h,x}| = 0.16$, $|g_h(\theta=80°)| = \sqrt{g_{h,z}^2 \cos^2\theta + g_{h,x}^2 \sin^2\theta} = 0.19$. Hence we obtain $|g_{h,z}| = 0.66$. Extrapolation of the dependence $\hbar\omega_h(B)$ to zero field does not show any offset for the valence band hole, just as it was the case in the analogous structure without contacts [13]. This indicates that the exchange interaction of the valence band holes with the ferromagnet is negligible, which is in agreement with the data in Fig. 2b (stars). Also in the magnetic field dependence of the electron Zeeman splitting $\hbar\omega_e(B)$ no offset is observed (not shown). Thus, neither the valence band hole, nor the electron experience an exchange interaction with the ferromagnet.

## S2. FM proximity effect at forward bias

Here we show that the decrease of the FM proximity effect for forward bias in the cw as well as time-resolved PL and the SRFS experiments is related with the appearance of valence band holes in the quantum well. The situation is illustrated in Fig. S2. Figure S2a shows the band structure diagram in equilibrium ($U = 0$). The holes fill the acceptor states, while holes in the valence band are absent. Application of a positive voltage $U > +0.5$ V flattens the bands, so that a fraction of holes moves from interface states to the valence band in the QW (Fig. S2b). A further increase of $U$ does not lead to band bending because the voltage is distributed throughout the structure, and the electric field in the QW region is small in accordance with the PL spectra. This picture is confirmed by all our experiments, as we will now discuss. SFRS gives the clearest proof of the appearance of valence band holes in the QW for forward bias.

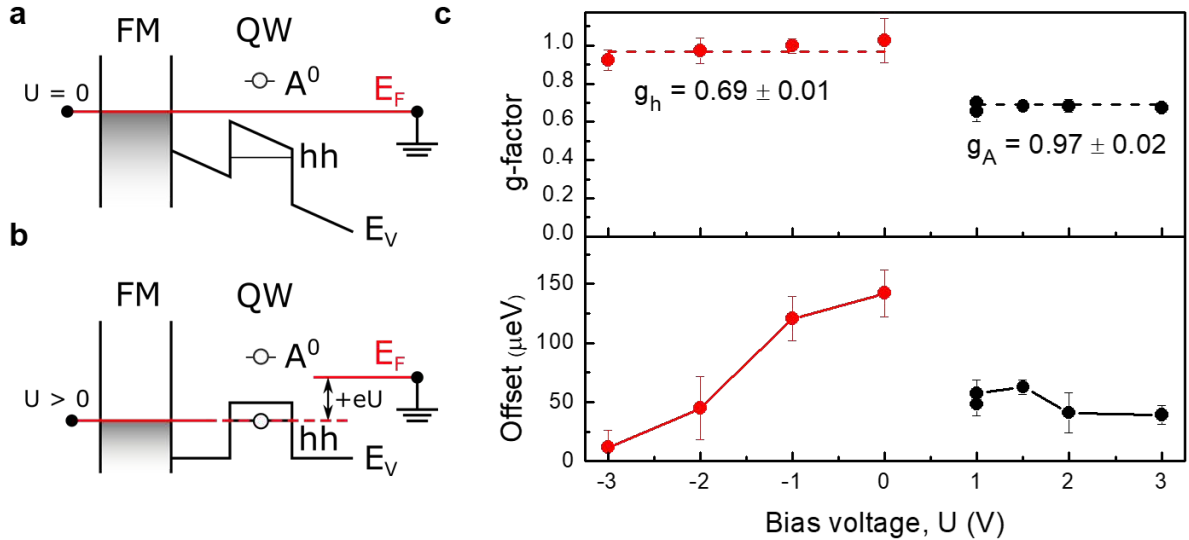

Fig. S2. Section of the hybrid structure band diagram in equilibrium (a) and for forward bias (flat band conditions) (b). The Fermi level $E_F$, the energy of the neutral acceptor $A^0$, and the lowest QW energy level of the heavy holes in the valence band (hh) are shown schematically. (c) Hole $g$-factor (top) and offset energy (bottom) dependences on the bias voltage, measured by SFRS.

The magnetic field is tilted by 20° with respect to the sample growth $z$-axis, $T = 2$ K. The dashed lines are guidelines for the eye.

In addition to the offset, SFRS gives the hole $g$-factor (of the valence band hole or of the hole bound to acceptor). When the $A^0X$ transition is excited, a spin flip of the acceptor hole can occur. From the slope of the linear fit to the magnetic field dependence of the Raman shift for reverse biases (see Fig. 4c) one calculates the $g$-factor of the hole bound to an acceptor to be $|g_A| = 0.97 \pm 0.02$. However, for forward bias the hole $g$-factor changes to 0.69 (Figure S2c, top), and the offset value decreases from 140 to 50 µeV (bottom). Interestingly, the $g$-factor of $0.69 \pm 0.01$ is close to the value of the longitudinal $g$-factor of the valence band hole $|g_{h,z}| = 0.66$, as determined from the pump-probe measurements (see Fig. S1a). These results indicate the contribution of valence band holes to SFRS at $U > +0.5$ V, in addition to holes bound to acceptors in accordance with Fig. S2a. The presence of holes in the valence band enables one to excite positively charged excitons ($X^+$ trion), quasiparticles consisting of two band holes and an electron, in the SFRS experiment. The binding energies of the $A^0X$ complex and the $X^+$ trion are close to each other. Under resonant excitation of the $X^+$ trion, the spin of the resident valence band hole flips according to the same mechanisms as in the acceptor hole spin flip in Fig. S3. Therefore, the appearance of a significant amount of valence band holes leads not only to an effective decrease of the Zeeman splitting offset (the spin flip lines of the band hole and the acceptor hole overlap strongly), but also to a change in the slope of the magnetic field dependence of the Stokes shift. Since the $p$-$d$ exchange with the band hole is much smaller than that with the hole bound to an acceptor, the measured $\Delta_{pd}$ value decreases.

The presence of valence band holes also explains the sharp decrease of the FM proximity effect with increasing voltage, followed by saturation for $U > +0.5$ V, as observed in the polarization-resolved cw (Fig. 2d) and time-resolved PL (Fig. 3b) on the e-$A^0$ line. The corresponding Eqs. (1.1) and (2.1) for the polarization of the holes bound to acceptors (the e-$A^0$ line) are valid only in the absence of valence band holes. Scattering with a mutual spin-flip (flip-flop transition) of the holes on neutral acceptors will lead to an averaging of the hole polarization in different orbital states. This mechanism of averaging is well known for conduction band electrons that scatter on electrons bound to neutral donors [S2]. The valence band holes interact weakly with the FM, and therefore the average polarization of the hole spin system will be reduced due to the spin exchange.

Thus, the results of all four experiments (cw and time-resolved PL, pump-probe Kerr rotation, SFRS) are consistent with each other, if we take into account the appearance of valence band holes in the QW (in addition to $A^0$) for forward bias. The valence band holes do not appear due to electrical injection, because the effects are saturated and do not scale exponentially with voltage like the correspondingly increasing current. Photo-injection of the holes can be excluded as well, since the SFRS measurements were carried out under resonant excitation below the interband and exciton absorption, and the behavior of the polarization dependences (see Figs. 2d and 3b) does not change with light intensity (at excitation power levels $\leq 5$ W/cm$^2$). The most probable reason for the appearance of valence band holes is charging of the quantum well in darkness, as illustrated in Fig. S2a, b.

**S3. Spin-flip Raman scattering on neutral acceptor**

Here, we consider the main physical processes of the spin flip of a hole bound to an acceptor. We use Faraday geometry (direction of excitation and scattered light are parallel to the magnetic field) while the sample is tilted by a small angle $\theta \ll 1$ ($\cos\theta \approx 1$) between the magnetic field direction and the $z$-axis. This relaxes the selection rules which are dictated by angular momentum conservation due to mixing of the electron states with spin projections of +1/2 and -1/2 onto the z-axis (↑ and ↓, respectively, as indicated in Fig. S3). The mixing parameter is given by $\beta \approx \theta/2$. In all cases the resonant intermediate (virtual) state is given by the exciton complex bound to a neutral acceptor ($A^0X$).

Figure S3a shows the double spin flip (DSF) scattering process [14]. This scattering involves a spin flip of the electron in the photoexcited exciton which is accompanied by emission of an acoustic phonon whose energy $\hbar\omega_q$ is equal to the Zeeman splitting $\mu_B|g_e|B$ of the conduction band electron in the $A^0X$ complex. In the initial state, a $\sigma^-$ photon comes in with energy $\hbar\omega_1$ (tuned to the $A^0X$ optical transition) and the acceptor has an angular momentum projection +3/2 as indicated by ⇑ in Fig. S3. In the final state, there are a $\sigma^+$ photon with energy $\hbar\omega_2$, the phonon with energy $\hbar\omega_q$, and an acceptor with momentum projection -3/2. It follows from energy conservation that the Stokes shift for the double spin flip $\Delta_S^{DSF} = \hbar\omega_1 - \hbar\omega_2 = \mu_B(|g_e| + |g_A|)B$ is determined both by the g-factor of the electron ($g_e$) and of the acceptor-bound hole ($g_A$). In the presence of p-d exchange interaction, there is an additional contribution $-\Delta_{pd}$ to the Stokes shift, so that we get the Eq. 3.1.

The single spin flip (SSF) scattering process is shown in Fig.S3b. Here, a phonon assisted spin flip of the conduction band electron in the excited state is not required. In presence of the p-d exchange interaction, the Stokes shift for the SSF is given by Eq. 3.2, which is determined only by the g-factor of the hole on the acceptor and by the exchange splitting. This process was observed in Ref. [13], which allowed us to directly measure $\Delta_{pd}$. Both the DSF and SSF mechanisms occur in the structure studied here (see Fig. 4a, lines "e+h" and "h", respectively). The change in the angular momentum of the photons by 2 quanta is analyzed using a crossed circular polarizer and an analyzer for both mechanisms.

Finally, there is a third spin-flip process, which corresponds to the single electron spin flip accompanied by emission of an acoustic phonon with energy $\hbar\omega_q = \mu_B |g_e| B$ in the $A^0X$ complex. It becomes allowed due to the *B*-field induced mixing of the states +1/2 and -1/2. This mechanism corresponds to the "e" line in the SFRS spectrum in Fig. 4a.

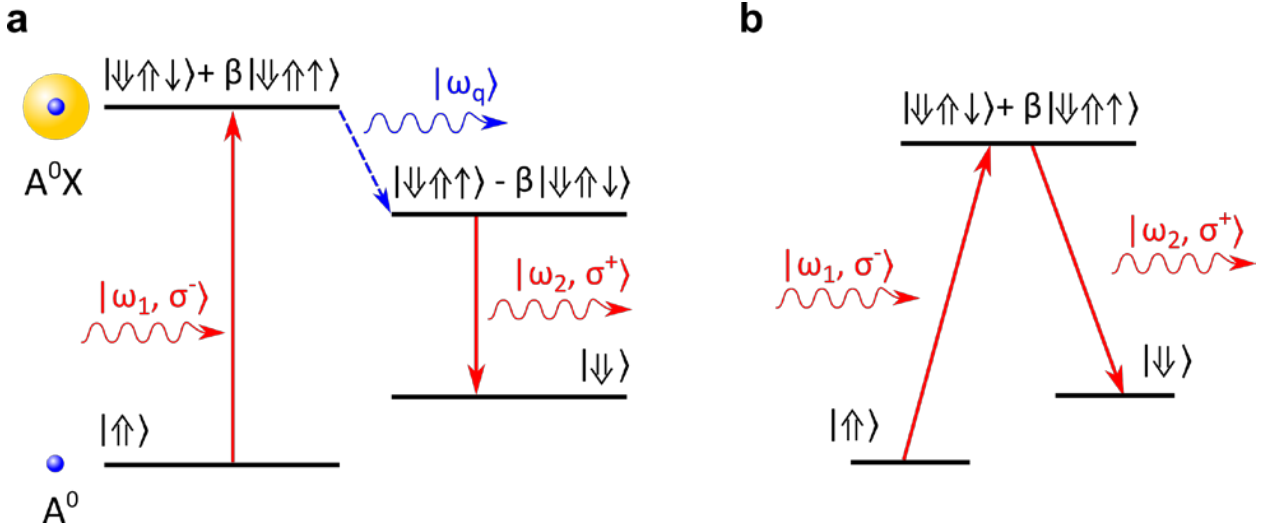

Fig.S3. Energy diagram of SFRS in magnetic field $B > \Delta_{pd}/\mu_B |g_A|$ (a) Scheme of the double spin flip (DSF). Here, ⇑, ⇓ indicate the angular momentum projections of the acceptor $J_z$ = +3/2, -3/2 and ↑, ↓ the electron spin projections $s_z$ = +1/2, -1/2 onto the *z*-axis. (b) Scheme of the single spin flip (SSF).

---

S1 Sirenko, A. A., Ruf, T., Cardona, M., Yakovlev, D. R., Ossau, W., Waag, A., and Landwehr, G. Electron and hole g factors measured by spin-flip Raman scattering in CdTe/CdMgTe single quantum wells. *Phys. Rev. B* **56**, 2114 (1997).

S2. Paget, D. Optical detection of NMR in high-purity GaAs under optical pumping: Efficient spin-exchange averaging between electronic states. *Phys. Rev. B* **24**, 3776 (1981).